\documentclass[aps,showpacs,prl,twocolumn]{revtex4}
\usepackage{amssymb,amsfonts,amsmath}
\makeatletter
\renewcommand{\@makefntext}[1]{\parindent=1em\noindent\hbox to 1.8em{\hss$^{\@thefnmark}$}#1}
\renewcommand{\@footnotemark}{\hbox{\mathsurround=0pt$^{\@thefnmark}$}}
\newcommand{\ftnote}[2]{\footnotemark[#1]\footnotetext[#1]{#2}}
\makeatother

\begin{document}

\title{Chiral symmetry and the string description of excited hadrons}

\author{L. Ya. Glozman}
\affiliation{Institute for Physics, Theoretical Physics Branch, University of Graz, Universit\"atsplatz 5, A-8010 Graz, Austria}

\author{A. V. Nefediev}
\affiliation{Institute of Theoretical and Experimental Physics, 117218, B.Cheremushkinskaya 25, Moscow, Russia}

\begin{abstract}
A large symmetry group is perhaps experimentally observed in excited hadrons which includes the chiral group $U(2)_L \times U(2)_R$ as a subgroup.
To possess this large symmetry a dynamical model for excited hadrons, presumably a string model, should explain formation of chiral 
multiplets
and, at the same time, predict coinciding slopes of the angular and radial Regge trajectories. This is possible only if both the dynamics of
the string and the chirality of the quarks at the ends of the string are considered together.
We construct a model--independent unitary transformation from the relativistic
chiral basis to the $\{{}^{2S+1}L_J\}$ basis, 
commonly used in hadronic phenomenology as well as in the string models, and demonstrate that a hadron belonging to the given chiral 
representation is a fixed superposition of the basis vectors with different $L$'s. 
Thus the description of highly excited hadron in terms of a fixed $L$ is not compatible with chiral symmetry and 
has to be disregarded in favour of the description in terms of the total hadron spin $J$. Therefore, 
dynamics of the string must deliver the principal quantum number $\sim n+J$,
in order chiral multiplets with different spins to become degenerate, 
as required by the large symmetry group. 
\end{abstract}

\maketitle

Recent remarkable experimental results on highly excited mesons
\cite{BUGG1,BUGG2} from the $p\bar{p}$ annihilation at LEAR (CERN)
reveal different kinds of symmetries of highly excited $n\bar{n}$
mesons\ftnote{1}{These results have to be confirmed by an independent experiment.
In particular, some of still missing states may be found.}: (i) equality of slopes of the angular and radial Regge
trajectories \cite{ANI,BUGG2}, (ii) effective restoration of the chiral $SU(2)_L \times SU(2)_R$
and $U(1)_A$ symmetries \cite{G1,G2} (see Ref.~\cite{G3} for a review),
(iii) clustering of many states with different spins, parities, and isospins around definite energies 
\cite{G4,BUGG2,AF,G3} (the latter property is obvious from
Fig.~2 of Ref.~\cite{G3}). This clustering
implies that perhaps a large symmetry group is observed which includes
chiral symmetries as subgroups\ftnote{2}{In this respect it is important to find
still missing chiral multiplets for the high--spin states at the levels
$M \sim 2.3$ GeV,  $M \sim 2$ GeV and, possibly, at $M \sim 1.7$ GeV.}.
Similar symmetries are also seen
in excited baryon spectra \cite{G5,G4,G3}. These results suggest
that physics, and in particular mass generation mechanisms, are essentially
different in the low- and high--lying states. The fundamental underlying reason for this difference comes from the fact that
quantum loop effects, which govern the properties of the low--lying states, are suppressed in high--lying hadrons, 
so that the dynamics becomes semi-classical \cite{G6,GN,G3}. Certainly understanding
this large symmetry group is one of the key problems to approach
QCD in the infrared, in particular in what concerns confinement and its interrelation
with chiral symmetry breaking.

Both the chiral restoration and asymptotically linear angular and radial
Regge trajectories are reproduced \cite{WG} within the only
known exactly solvable chirally symmetric and confining model (Generalised Nambu--Jona-Lasinio (GNJL) model) \cite{GNJL,Lisbon}. In
this model the only gluonic interaction is the linear Coulomb-like
confining potential. Then chiral symmetry breaking is obtained
from the Schwinger--Dyson (mass--gap) equation, and the meson
spectrum results from a Bethe--Salpeter equation \cite{GNJL,Lisbon,f,NR1}. 
While chiral symmetry restoration in excited hadrons is naturally explained in the framework of this model \cite{chisym} and
all possible highly excited states with the same $J$ and $n$ fall into
$[(0,1/2)+(1/2,0)] \times  [(0,1/2)+(1/2,0)]$ chiral representation 
(which combines all possible chiral multiplets with the same $J$) and become
approximately degenerate \cite{WG}, the degeneracy of states with different $J$'s is
absent, because the slopes of the radial and angular Regge trajectories are
different. It is not possible to obtain simultaneously chiral 
symmetry restoration and equal slopes of angular and radial Regge trajectories
within any equal--time relativistic {\it potential} description \cite{B}.

Equal slopes of the radial and angular Regge trajectories is a
generic property of the Veneziano dual amplitude \cite{V} and of
the open bosonic Nambu--Goto string \cite{NG}. Quarks and the notion of chiral
symmetry are absent in this pre--QCD approach, however. 

There are modern attempts to model QCD within the AdS/QCD approach \cite{ADS}, based on the
ideas of Maldacena on duality between the conformal supersymmetric field 
theory and the string theory in AdS. QCD in the infrared is a highly
nonconformal theory, however, (which is evidenced already by the presence
of the scale --- the Regge slope), and such a modeling is based in fact on
ad hoc and uncontrolled deformations in the infrared which simulate
confinement and break conformal symmetry. The interrelations
between chiral symmetry breaking and confinement, as demanded
in QCD by the 't Hooft anomaly matching conditions \cite{AN} and by the
Coleman--Witten theorem \cite{CW},
are not clear in these models and it is possible to achieve, at least
in some AdS/QCD models, confinement without manifest chiral symmetry 
breaking in the vacuum. It is possible to choose the infrared boundary
in this approach in such a way that the radial and angular Regge
slopes coincide \cite{SON}, however chiral multiplets of excited hadrons do not show up.

In this note we address the question whether it is possible
or not to reconcile a view of an excited hadron as an open bosonic
string (electric flux tube) with quarks at the ends
with the approximately restored chiral symmetry in this hadron.
It is generally assumed within such a picture that quarks serve mainly as sources of the colour--electric field.
The energy of the system is stored in the gluonic electric string (flux tube) which generates an effective interquark interaction 
traditionally described in terms of the string orbital angular momentum $L$ (for example, in a specific
approach of this kind \cite{ft2} the resulting interquark interaction is strongly angular momentum ($L$) dependent, 
is nonlocal and does not amount to a plain potential \cite{regge}). 
The spins of the quarks, irrelevant for the dynamics of the string, 
are used then to obtain the total spin of the hadron within the LS-coupling
scheme, $\vec J = \vec L + \vec S$, i.e.
 $J=L-1,L,L+1$  
(the dependence on the quark spins orientation enters for low--lying states only through corrections due to spin--dependent terms in 
the effective interquark 
potential). Then the energy of the hadron is determined, like in the Nambu--Goto string, by
the orbital angular momentum $L$ of the flux tube, that describes its rotational motion, and by the radial quantum number $n$, that
describes the vibrational motion of the string (we ignore here possible transverse excitations of the flux tube and hence do
not consider hybrids in this paper). 

Such a picture has obtained a certain support in lattice simulations with the static sources \cite{BA},
where at large separation between the sources only electric flux tube is seen,
that can be approximated by a linear potential. The observed flux tube
is highly nondynamical, however, because its ends are static. 
Both  the Nambu--Goto string as well as  the static lattice simulations do not appeal to chiral degrees of freedom which are present
in the light quark hadrons. 

There are  notorious experimental and theoretical reasons to believe
that chiral symmetry is highly relevant and is actually  approximately manifest in the highly excited hadrons.
Then the question arises whether the popular nonrelativistic
$\{{}^{2S+1}L_J\}$--inspired classification scheme
and dynamical pictures based upon it are adequate for highly excited hadrons?
In other words, we put in question the very possibility to describe highly excited hadrons in terms of the quantum numbers $n$ and $L$, for 
the latter do not comply with chiral symmetry. Hence, in a dynamically generated hadron consisting of the quarks at the ends of the string,
the role of the quarks and in particular of their spin orientations is highly nontrivial, contrary to 
the traditional belief. Below we prove this statement. As a byproduct it is then clear
that the nonrelativistic $\{{}^{2S+1}L_J\}$ classification,
traditionally used in the quark picture and somewhat justified in the heavy quark systems, makes no sense in the
highly excited light--quark hadrons with the approximately restored chiral
symmetry, because $L$ is not a separately
conserved quantity. The chiral basis is to be used instead.

\begin{table}[t]
\caption{The complete set of $q\bar{q}$ states classified according to the chiral basis. 
The sign $\leftrightarrow$ indicates the states belonging to the same representation which become approximately degenerate.}
\begin{ruledtabular}
\begin{tabular}{cccc}
$R$&$J=0$&$J=1,3,\ldots$&$J=2,4,\ldots$\\
\hline
$(0,0)$&\bf ---&$0J^{++} \leftrightarrow 0J^{--}$&$0J^{--}  \leftrightarrow 0J^{++}$\\
$(1/2,1/2)_a$&$1J^{-+}\leftrightarrow 0J^{++}$&$1J^{+-} \leftrightarrow 0J^{--}$&$1J^{-+} \leftrightarrow 0J^{++}$\\
$(1/2,1/2)_b$&$1J^{++}\leftrightarrow 0J^{-+}$&$1J^{--} \leftrightarrow 0J^{+-}$&$1J^{++} \leftrightarrow 0J^{-+}$\\
$(0,1) \oplus (1,0)$&\bf ---&$1J^{--} \leftrightarrow 1J^{++}$&$1J^{++} \leftrightarrow 1J^{--}$\\
\end{tabular}\label{t1}
\end{ruledtabular}
\end{table}

Before diving into technicalities we briefly outline the content of the
proof. Since both the basis of chiral multiplets and the $\{{}^{2S+1}L_J\}$ basis are
complete ones, then we can construct a unitary transformation from
one basis to the other, similar to the transformation from
the two--body relativistic helicity basis to the nonrelativistic one \cite{LL}.
This transformation is a pure mathematical enterprise and is completely
model--independent. While the derivation of this unitary transformation is
rather straightforward and simple, it has quite interesting physical implications
directly relevant to hadron modelling. It turns out that
a given chiral state that belongs to some of
the chiral multiplets is a fixed superposition of two $\{{}^{2S+1}L_J\}$ basis
states with different $L$'s. Hence if chiral symmetry is approximately
restored in a given excited hadron, its description in terms of a
fixed $L$ is impossible. The energy of the dynamical flux tube cannot be 
determined  solely by its orbital angular momentum  $L$ and can depend
generally only on $J$ and on the chiral index. 
Hence the role of the quark spin orientations 
is highly nontrivial for the energy of the dynamical flux tube. 

Chiral multiplets of the $SU(2)_L \times SU(2)_R$ group for $q\bar{q}$ mesons were classified in Refs.~\cite{G2,G3}. They can be naturally
described in terms of the chiral basis with the basis vectors depending on the index of the chiral 
representation $R$ ($R=$ $(0,0)$, $(1/2,1/2)_a$, $(1/2,1/2)_b$, or $(0,1)+(1,0)$), on the spatial parity $P$, on the total spin $J$ and its
projection $M$, as well as on the isospin $I$ and its projection $i$.
To simplify notations, we omit the spin and isospin projections in the notation of the
chiral basis vectors, referring to them as to $|R;IJ^{PC}\rangle$, 
where for a neutral $q\bar{q}$ system the $C$--parity is related to the other quantum
numbers in the standard way \cite{LL}.
The chiral basis $\{R;IJ^{PC}\}$ is obviously consistent with the Poincar{\'e} invariance. Highly excited hadrons belonging 
to the same chiral representation $R$ but possessing opposite parities $P$ are approximately degenerate. Hence the chiral basis
is extremely convenient for describing hadron wave functions. In Table~\ref{t1} 
we give the complete set of $q\bar{q}$ states classified according to this chiral basis. 

The chiral basis vectors can be constructed as
\begin{equation}
|R;IJ^{PC}\rangle=\sum_{\lambda_q\lambda_{\bar q}}\sum_{i_q  i_{\bar q}}
\chi_{\lambda_q \lambda_{\bar q}}^{RPI}C_{\frac12 i_q\frac12 i_{\bar q}}^{Ii}
|i_q\rangle |i_{\bar q}\rangle|J\lambda_q\lambda_{\bar q}\rangle,
\label{bvec}
\end{equation}
where $i_q$ and $\lambda_q$ ($i_{\bar q}$ and $\lambda_{\bar q}$) are the quark (antiquark) isospin and 
helicity, respectively. 
The coefficients $\chi_{\lambda_q \lambda_{\bar q}}^{RPI}$ 
can be extracted from the explicit form of the basis vectors
given in Refs.~\cite{G2,G3}. They 
form a unitary transformation from the quark helicity basis to the 
chiral basis with definite parity in the state.
In Table~\ref{t2} we give these coefficients for various
chiral representations and quantum numbers. 
Notice that the chiral basis is closely related to the notion of the quark helicity as
helicity of the quark coincides with its chirality while helicity of the antiquark is just opposite to its chirality. In other words, 
one can refer to the states $|J\lambda_q\lambda_{\bar q}\rangle$ as to the states with the given chirality of the quarks.
Vectors $|J\lambda_q\lambda_{\bar q}\rangle$ do not possess definite parity --- states with a definite parity
can be constructed only as linear combinations of such vectors with opposite helicity.

To proceed we are
to build an explicit form of the vectors $|J\lambda_q\lambda_{\bar q}\rangle$ in terms of the single--quark 
states with the given helicity
and to establish the relation between these states and the basis vectors of the $\{{}^{2S+1}L_J\}$ scheme, that is to find the matrix 
$\langle J\lambda_q \lambda_{\bar q}|{}^{2S+1}L_J\rangle$, where $L$ is the orbital angular
momentum and $S$ is the total spin of two particles. Then, following Ref.~\cite{LL}, 
we use\ftnote{3}{We use the standard definition of the phase of the spherical functions, which
differs by the factor $i^L$ from the definition in Ref.~\cite{LL}}
\begin{equation}
|J\lambda_q\lambda_{\bar q}\rangle=D^{(J)}_{\lambda_q - \lambda_{\bar q}, M}(\vec n)\sqrt {\frac{2J+1}{4\pi}}
|\lambda_q\rangle  |-\lambda_{\bar q}\rangle,
\label{hel}
\end{equation}
where $D^{(J)}_{MM'}(\vec n)$ is the standard Wigner $D$--function describing rotation from the quantization axis to
the quark momentum direction $\vec n=\vec{p}/p$. Finally, after simple algebraic transformations, we find:
\begin{eqnarray}
\langle J\lambda_q \lambda_{\bar q}|{}^{2S+1}L_J\rangle&=&\sqrt{\frac{2L+1}{2J+1}}
C_{\frac12\lambda _q\frac12-\lambda_{\bar q}}^{S\Lambda}C_{L0S\Lambda}^{J\Lambda},\nonumber\\[-2mm]
\label{un}\\[-2mm]
\Lambda&=&\lambda _q-\lambda_{\bar q}.\nonumber
\end{eqnarray}

Combining Eqs.~(\ref{bvec})--(\ref{un}) together and using the property that
$\langle {}^{2S+1}L_J|J\lambda_q \lambda_{\bar q}\rangle=\langle J\lambda_q \lambda_{\bar q}|{}^{2S+1}L_J\rangle^*$, we arrive at:
\begin{eqnarray}
|R;IJ^{PC}\rangle&=&\sum_{LS}\sum_{\lambda_q\lambda_{\bar q}}\sum_{i_q  i_{\bar q}}
\chi_{\lambda_q \lambda_{\bar q}}^{RPI}
C_{\frac12  i_q  \frac12  i_{\bar q}}^{Ii}|i_q\rangle |i_{\bar q}\rangle\nonumber\\[-2mm]
\\[-2mm]
&\times&\sqrt{\frac{2L+1}{2J+1}}C_{\frac12\lambda _q\frac12-\lambda_{\bar q}}^{S\Lambda}
C_{L0S\Lambda}^{J\Lambda}|{}^{2S+1}L_J\rangle.\nonumber
\end{eqnarray}
Consequently, one ends up with a unitary transformation
from the basis vectors of the $SU(2)_L \times SU(2)_R$ chiral group with the
fixed $IJ^{PC}$ quantum numbers to the $\{I;{}^{2S+1}L_J\}$ basis:
\begin{eqnarray}
|R;IJ^{PC}\rangle&=&\sum_{LS}\sum_{\lambda_q\lambda_{\bar q}}\chi_{\lambda_q \lambda_{\bar q}}^{RPI}\nonumber\\[-2mm]
\label{5}\\[-2mm]
&\times&\sqrt{\frac{2L+1}{2J+1}}C_{\frac12\lambda _q\frac12-\lambda_{\bar q}}^{S\Lambda}
C_{L0S\Lambda}^{J\Lambda}|I;{}^{2S+1}L_J\rangle.\nonumber
\end{eqnarray}
Hence every state from the chiral basis is a fixed superposition
of allowed states in the $\{I;{}^{2S+1}L_J\}$ basis.
Note that in the sum above only those $L$ are allowed which satisfy
$P=(-1)^{L+1}$. 

The unitary transformation derived above is a model--independent result. It implies
stringent limitations on description of physical states with approximate
chiral symmetry. If chiral symmetry breaking in a given physical
state is only a small perturbation, 
this state is described by one of the basis vectors $|R;IJ^{PC}\rangle$ to a good accuracy.
For particular basis vectors the sum in Eq.~(\ref{5}) is restricted to only one fixed combination $\{L,S\}$ (still,
there is a summation over the quark helicities).
For example, for the axial--vector mesons $a_1$, described by the chiral vector $|(0,1)+(1,0);1~1^{++}\rangle$, 
the sum in $L$ and $S$ contains only one term, namely $|1;{}^3P_1\rangle$. Similarly,
the $h_1$ meson corresponds to the state $|(1/2,1/2)_b ;0 ~ 1^{+-}\rangle$, in the chiral basis, and to the state 
$|0;{}^1P_1\rangle$, in the $\{I;{}^{2S+1}L_J\}$ basis. However, there are two kinds of the $\rho$--mesons, given by two different
vectors, $|(0,1)+(1,0);1 ~ 1^{--}\rangle$ and $|(1/2,1/2)_b ;1 ~ 1^{--}\rangle$, in the chiral basis.
Each of these $\rho$-mesons is represented by the mutually orthogonal {\it fixed} superpositions of two different
partial waves:
\begin{eqnarray}
\displaystyle |(0,1)+(1,0);1 ~ 1^{--}\rangle&=&\sqrt{\frac23}|1;{}^3S_1\rangle+\sqrt{\frac13}|1;{}^3D_1\rangle,\nonumber\\
\displaystyle |(1/2,1/2)_b;1 ~ 1^{--}\rangle&=&\sqrt{\frac13}|1;{}^3S_1\rangle-\sqrt{\frac23}|1;{}^3D_1\rangle.\nonumber
\end{eqnarray}
Hence the description of the $\rho$--meson with approximately restored chiral symmetry is impossible
in terms of a fixed $L$. Obviously, this situation occurs for all mesons from Table~\ref{t1} for which
two different chiral representations can be assigned to the given $IJ^{PC}$ set. 
Consequently, for many mesons with $J>0$ there must be a duplication of the Regge trajectories, each of them can be uniquely specified by
a proper chiral index. Such a duplication of the Regge trajectories is indeed observed in the Crystal Barrel data \cite{BUGG1,BUGG2}.

One has to conclude that generally the description of the highly excited hadronic states with 
approximately restored chiral symmetry is impossible in terms of the fixed $L$. 
This simple result calls for a review of the thirty--year tradition in classification of the light--quark 
sector based on the naive quark model scheme\ftnote{4}{Note that the notion of the spin--orbit force is not
defined for the states with a definite chirality  since the chirality operator does not commute 
with the orbital angular momentum operator $L$.
This is directly relevant to the widely discussed ``problem" of missing spin--orbit
force \cite{G3,G4}.}.
A naive string picture with the fixed $L$ is to be revised either to include quark chiralities and thus to become 
compatible with chiral symmetry \cite{G3,G4}. A possible string description of the large experimental degeneracy of mesons
with different spins, isospins, and parities
would amount to demonstration that the string energy is determined only by the principal quantum
number and hence is independent on $|R;IJ^{PC}\rangle$. In particular, the large degeneracy seen at Fig.~2
of Ref.~\cite{G3} might be understood if, on top of the chiral symmetry restoration, a principal
quantum number $N=n+J$ existed. Note that the chiral restoration requires chiral multiplets
for the highest spin states, which are presently missing in Refs.~\cite{BUGG1,BUGG2}. Hence it is a very important
experimental task to find them or to reliably establish their absence\ftnote{5}{Such high--spin parity doublets
are well seen in the nucleon spectrum --- see Fig.~1 of Ref.~\cite{G3}}.
\begin{table}[t]
\caption{The complete set of nonzero coefficients 
$\chi_{\lambda_q \lambda_{\bar q}}^{RPI}$.}
\begin{ruledtabular}
\begin{tabular}{c}
$R=(0,0)\quad P=(-1)^J\quad I=0$\\[0.5mm]
$\chi_{\frac12 -\frac12}^{RPI}=
\chi_{-\frac12 \frac12}^{RPI}=
\displaystyle\frac{1}{\sqrt{2}}$\\[0.5mm]
\hline
$R=(0,0)\quad P=(-1)^{J+1}\quad I=0$\\[0.5mm]
$\chi_{\frac12 -\frac12}^{RPI}=
-\chi_{-\frac12 \frac12}^{RPI}=
\displaystyle\frac{1}{\sqrt{2}}$\\[0.5mm]
\hline
\hline
$R=(1/2,1/2)_a\quad P=(-1)^J\quad I=0$\\[0.5mm]
$\chi_{-\frac12 -\frac12}^{RPI}=
\chi_{\frac12 \frac12}^{RPI}=
\displaystyle\frac{1}{\sqrt{2}}$\\[0.5mm]
\hline
$R=(1/2,1/2)_a\quad P=(-1)^{J+1}\quad I=1$\\[0.5mm]
$\chi_{-\frac12 -\frac12}^{RPI}=
-\chi_{\frac12 \frac12}^{RPI}=
\displaystyle\frac{1}{\sqrt{2}}$\\[0.5mm]
\hline
\hline
$R=(1/2,1/2)_b\quad P=(-1)^J\quad I=1$\\[0.5mm]
$\chi_{-\frac12 -\frac12}^{RPI}=
\chi_{\frac12 \frac12}^{RPI}=
\displaystyle\frac{1}{\sqrt{2}}$\\[0.5mm]
\hline
$R=(1/2,1/2)_b\quad P=(-1)^{J+1}\quad I=0$\\[0.5mm]
$\chi_{-\frac12 -\frac12}^{RPI}=
-\chi_{\frac12 \frac12}^{RPI}=
\displaystyle\frac{1}{\sqrt{2}}$\\[0.5mm]
\hline
\hline
$R=(0,1)\oplus(1,0)\quad P=(-1)^J\quad I=1$\\[0.5mm]
$\chi_{\frac12 -\frac12}^{RPI}=
\chi_{-\frac12 \frac12}^{RPI}=
\displaystyle\frac{1}{\sqrt{2}}$\\[0.5mm]
\hline
$R=(0,1)\oplus(1,0)\quad P=(-1)^{J+1}\quad I=1$\\[0.5mm]
$\chi_{\frac12 -\frac12}^{RPI}=
-\chi_{-\frac12 \frac12}^{RPI}=
\displaystyle\frac{1}{\sqrt{2}}$\\[0.5mm]
%\hline
\hline
\end{tabular}\label{t2}
\end{ruledtabular}
\end{table}

\acknowledgments
L.Ya.G. is grateful to Tom Cohen for useful comments and
acknowledges support of the Austrian Science Fund through grant
P19168-N16.
Work of A.N. was supported by the Federal Agency for Atomic
Energy of Russian Federation, by the grants RFFI-05-02-04012-NNIOa, DFG-436 RUS
113/820/0-1(R), NSh-843.2006.2, and PTDC/FIS/70843/2006-Fi\-si\-ca, as well as
by the Federal Programme of the Russian Ministry of
Industry, Science, and Technology No. 40.052.1.1.1112.

\end{document}